\documentclass[a4paper,twoside]{article}

\usepackage{epsfig}
\usepackage{subcaption}
\usepackage{calc}
\usepackage{amssymb}
\usepackage{amstext}
\usepackage{amsmath}
\usepackage{amsthm}
\usepackage{multicol}
\usepackage{pslatex}
\usepackage{apalike}
\usepackage{algorithm2e}
\usepackage[bottom]{footmisc}
\usepackage{SCITEPRESS}     
\usepackage{todonotes}
\usepackage{soul}
\usepackage{arydshln}

\usepackage{array}

\definecolor{gray50}{gray}{.5}
\definecolor{gray40}{gray}{.6}
\definecolor{gray30}{gray}{.7}
\definecolor{gray20}{gray}{.8}
\definecolor{gray10}{gray}{.9}
\definecolor{gray05}{gray}{.95}

\usepackage{hyperref}
\hypersetup{
    colorlinks=true,
    linkcolor=black,
    filecolor=black,      
    urlcolor=black,
    citecolor=black
    }

\begin{document}

\title{Service Weaver: A Promising Direction for Cloud-native Systems?}

\author{\authorname{Jacoby Johnson\sup{1}, Subash Kharel\sup{1} Alan Mannamplackal\sup{1}, Amr S. Abdelfattah\sup{1}, and Tomas Cerny\sup{2}}
\affiliation{\sup{1} Computer Science of Baylor University, Waco, Texas, United States}
\affiliation{\sup{2} Systems and Industrial Engineering, University of Arizona, Tucson, Arizona, United States}
\email{amr\_elsayed1@baylor.edu \& tcerny@arizona.edu}
}



\keywords{Service Weaver, Cloud-native Systems, Microservices, Modular Binary, Cloud Computing}

\abstract{Cloud-native and microservice architectures have taken over the development world by storm. While being incredibly scalable and resilient, microservice architectures also come at the cost of increased overhead to build and maintain. Google's Service Weaver aims to simplify the complexities associated with implementing cloud-native systems by introducing the concept of a single modular binary composed of agent-like components, thereby abstracting away the microservice architecture notion of individual services. While Service Weaver presents a promising approach to streamline the development of cloud-native applications and addresses nearly all significant aspects of conventional cloud-native systems, there are existing tradeoffs affecting the overall functionality of the system. Notably, Service Weaver's straightforward implementation and deployment of components alleviate the overhead of constructing a complex microservice architecture. However, it is important to acknowledge that certain features, including separate code bases, routing mechanisms, resiliency, and security, are presently lacking in the framework.}



\onecolumn \maketitle \normalsize \setcounter{footnote}{0} \vfill

\section{\uppercase{Introduction}}
\label{sec:introduction}

Cloud-native systems and the development of microservice architecture have taken center stage in the realm of distributed systems software. While microservices undeniably enhance functionality, offering improved availability and scalability, they introduce heightened system complexity and pose various challenges~\cite{abdelfattah2023roadmap,cerny2022microservicereview}. Beyond managing business logic, a microservice architecture must address inter-service communications, necessitating the implementation of service registry, discovery, and routing mechanisms. Additionally, the deployment process presents its own set of challenges, requiring careful configuration management for swift and reliable deployment to uphold system scalability~\cite{abdelfattah2023roadmap,bushong2021microservice}.

The potential for simplifying the intricate processes of creating, deploying, and maintaining microservices is indeed promising. In case of the complexities involved in managing such challenges could be abstracted for practitioners, the tasks associated with creating, deploying, and maintaining microservices would undergo a notable simplification. This abstraction could potentially streamline the development workflow, enhance efficiency, and empower developers to focus more on core application logic, fostering a more seamless and productive development experience.



Service Weaver~\cite{serviceweaver_hotos23,serviceweaver_dev} emerges to serve in this direction as a programming framework for writing and deploying cloud applications. It enables practitioners to write their program in a modular binary and then that binary can be deployed as a microservice through a lightweight configuration file. This offers developers flexibility for development and testing, whether it's done locally or in a cloud environment.


At first glance it serves as a focal point of exploration, prompting compelling questions about its capabilities and potential. At a high level, this could be revolutionary for cloud-native systems; however, perfect solutions simply don’t exist and an in-depth look at the benefits and the drawbacks of implementing a new technology must always be considered. This paper aims to delve into the question of: "Is Service Weaver capable of covering cloud-native components while maintaining the full functionality of a microservice architecture?" Unraveling this question will guide our investigation into the unique attributes and functionalities of the Service Weaver in the context of modern cloud-native architectures.





The rest of the paper is organized as follows, Section 2 stated the research questions. A brief background into existing cloud-native components is shared in Section 3. The heart of this paper, in Section 4, is dedicated to exploring Service Weaver capabilities and its potential application. In Section 5 the research questions were addressed, and in Section 6 the conclusion was drawn. While Service Weaver is a new technology, the authors believe it to be potentially impactful and could see it having a role in shaping the future of cloud-native systems.



\section{\uppercase{Research Questions}}


Service Weaver, with its modular binary approach, proposes a solution to simplify the microservices model of development. However, the efficiency of Service Weaver in meeting the core requirements of cloud-native systems, such as scalability, flexibility, and resilience, while dealing with the complexities of traditional microservices architectures, remains an open question. This question fuels this research to position the capabilities of Service Weaver within cloud-native context, assessing its suitability for building effective and robust cloud-native systems.

\vspace{-0.5em}
\subsubsection*{RQ$_{1}$: What advantages does writing your application as a modular binary offer in comparison to using Microservice architecture?}
\vspace{-0.5em}
 The modular binary methodology involves consolidating all components into a single executable file, a concept facilitated by the Service Weaver. This approach raises questions about the potential benefits in terms of reduced complexity and the adaptability to deploy modular components as microservices. This question navigates the modular approach's challenges such as network latency, debugging, testing, and resource management which are often associated with microservice architecture.
\vspace{-0.5em}
\subsubsection*{RQ$_{2}$: Is the Service Weaver capable of effectively covering cloud-native components?}
\vspace{-0.5em}
 Cloud-native components, known for enabling scalability, flexibility, and resilience in cloud computing environments, present a crucial consideration for modern applications. This question revolves around Service Weaver's ability to streamline application development, offering choices between a modular monolith and microservice-based systems development and deployment. It addresses the alignment of Service Weaver with the essential features required for cloud-native components, including separation of concerns, scalability, resiliency management, and deployment flexibility behavior across diverse deployment environments.


\section{Background}

Cloud-native and microservice architectures are centered around the creation of scalable, flexible, and resilient applications. To achieve these attributes, various components are introduced to assist practitioners in their development endeavors. The stack encompasses infrastructure, provisioning, runtime, orchestration, management, application definition, and observability layers~\cite{abdelfattah2023roadmap,aws_cloud_native}.

Microservices, foundational to cloud-native applications, are designed as small, interdependent services that break down functionalities into smaller, independent units to enhance agility and resource efficiency~\cite{abdelfattah2023roadmap,aws_cloud_native}. These microservices communicate with each other through network protocols and message passing to accomplish tasks. The scalability of cloud-native systems underscores the significance of the Service Discovery component, which dynamically locates the network locations of microservices, facilitating their interaction in distributed environments. This ensures seamless communication among services, adapting to changes in their network locations or instances~\cite{springmicroservices}. Additionally, the API Gateway component functions as a critical component in managing incoming requests to microservices, serving as an entry point for clients. It routes requests, handles load balancing, and may implement security measures such as authentication and rate limiting~\cite{springmicroservices}. Research leverages the API Gateway component to address recurring communication challenges with the client side~\cite{fillinggapscloser23}.


On the other hand, various dependencies are present in microservices architecture components, ranging from explicit inter-service calls to implicit data model sharing between microservices~\cite{abdelfattah2023microservice}. As the system evolves, managing and scaling these dependencies becomes complex, leading to the emergence of multiple bad practices (i.e., anti-patterns)~\cite{cerny2023catalog} in microservice implementation. Consequently, effectively managing dependencies poses challenges in maintaining the quality and health of the architecture.

\section{Comparing Modular and Microservice Architectures}

In this section, we delve into a comparative discussion of Modular Architecture and Microservice Architecture, two prevalent paradigms in the realm of software design. Each approach presents unique advantages and challenges, influencing how practitioners structure and organize their applications. 

\subsection{Modular Architecture}
Modular architecture emphasizes \textit{strong encapsulation} and \textit{well-defined interfaces}. This approach hides implementation details within components, leading to low coupling between different parts. It allows different teams to work independently on separate components, which then come together in a \textit{single deployment unit}. This fosters parallel development while maintaining overall system coherence. 

Modular architecture's advantage lies in its balance between monolithic and microservice architectures. It provides the isolation and independence of microservices while maintaining the simplicity and ease of development found in monolithic systems. This approach reduces the overhead for developers, as the system is more straightforward to understand and maintain, and it avoids the network complexity often associated with microservices. Modular architecture can be particularly beneficial for small to medium-sized teams or applications where full-scale microservices might be overkill~\cite{oreilly_modular_microservices}.
Additionally, this architecture fosters the \textit{Improved Maintainability}. With clear boundaries and well-defined responsibilities among modules, it simplifies bug fixing and adding new features. This leads to a better understanding of the codebase and reduces time and costs for maintenance tasks~\cite{appmaster_modular_advantages}.


\subsection{Microservice Architecture}
Microservice architecture, while offering \textit{scalability and flexibility}, introduces \textit{significant operational complexity}. It allows teams to \textit{work and scale independently}, as each microservice can be managed and scaled as a separate entity.

The primary advantage of microservice architecture is its ability to scale components independently and its resilience. Each service can be deployed, scaled, and updated without impacting other services, making the system more robust against failures. This architecture is ideal for large, complex applications where different services require different technologies and scalability needs. It facilitates continuous delivery and deployment, allowing businesses to respond rapidly to market changes or customer demands~\cite{oreilly_modular_microservices}.

Additional advantages of this architecture include \textit{Enhanced Team Productivity and Specialization}, allowing focused teams to develop and maintain specific services. This develops ownership and expertise.
\textit{Agility in Deployments} is enhanced, as services evolve and deploy independently, speeding up releases and reducing coordination risks~\cite{atlassian_microservices_advantages}.

\subsection{Clarification Example}


Illustrating the architectural structures of both modular and microservices frameworks, consider an example of an online retail application. This application encompasses the following functionalities: inventory management, order processing, customer account management, payment processing, shipping coordination, and search processing. Together, these functionalities seamlessly orchestrate the entire shopping experience, spanning from product selection to order fulfillment.

\begin{figure}[ht]
\centering
\vspace{-1em}
\includegraphics[width=0.4\textwidth]{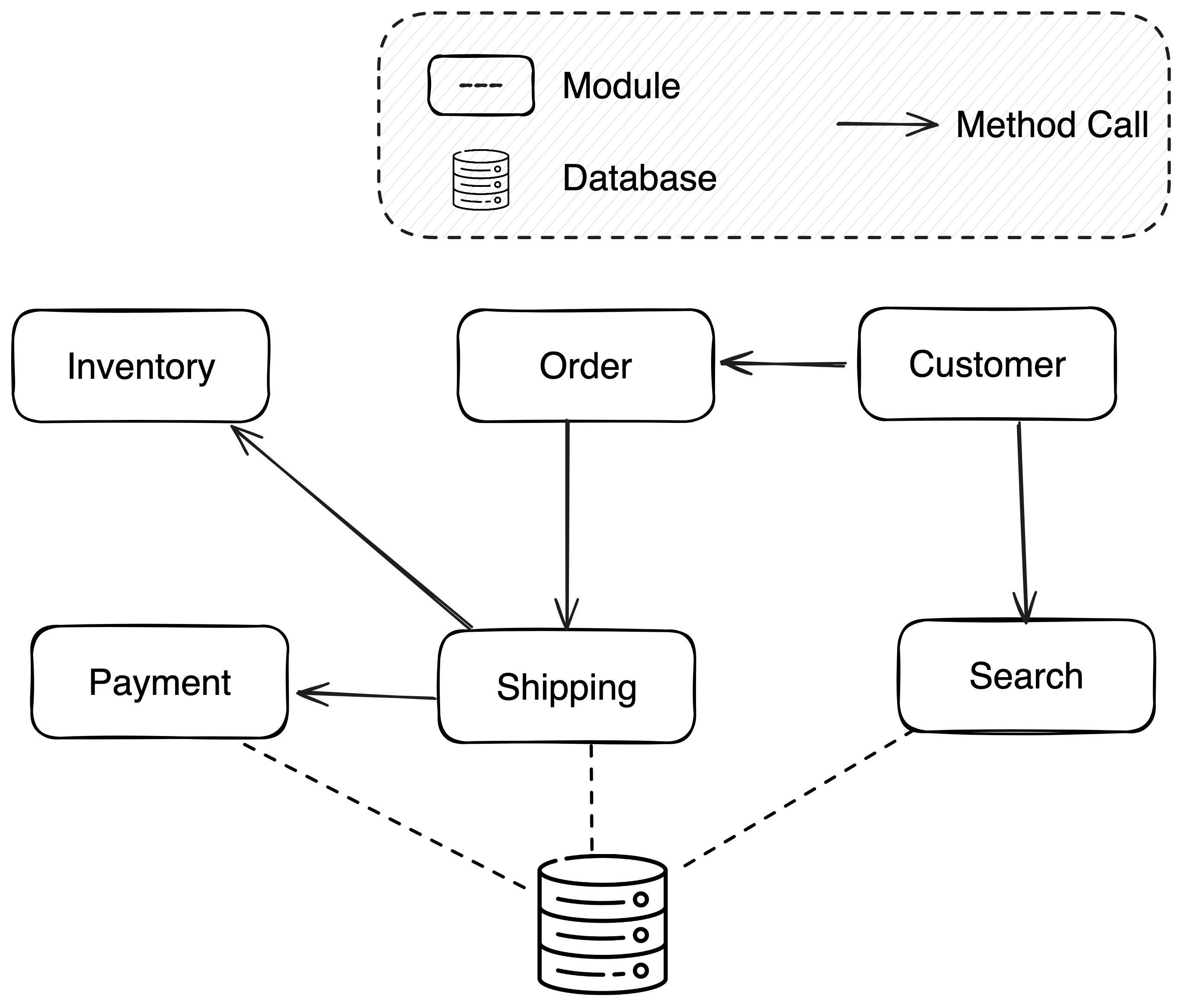}
\caption{Modular Architecture of a Retail Application}
\label{fig:modular_architecture}
\end{figure}

In a \textbf{modular architecture}, as depicted in Figure~\ref{fig:modular_architecture}, these functionalities are developed as distinct modules but within a single application. This means that while inventory management, order processing, and other services are separate in terms of their code and functions, they coexist within the same application boundary and share common resources like a central database.

\begin{figure}[ht]
\centering
\vspace{-0.5em}
\includegraphics[width=0.4\textwidth]{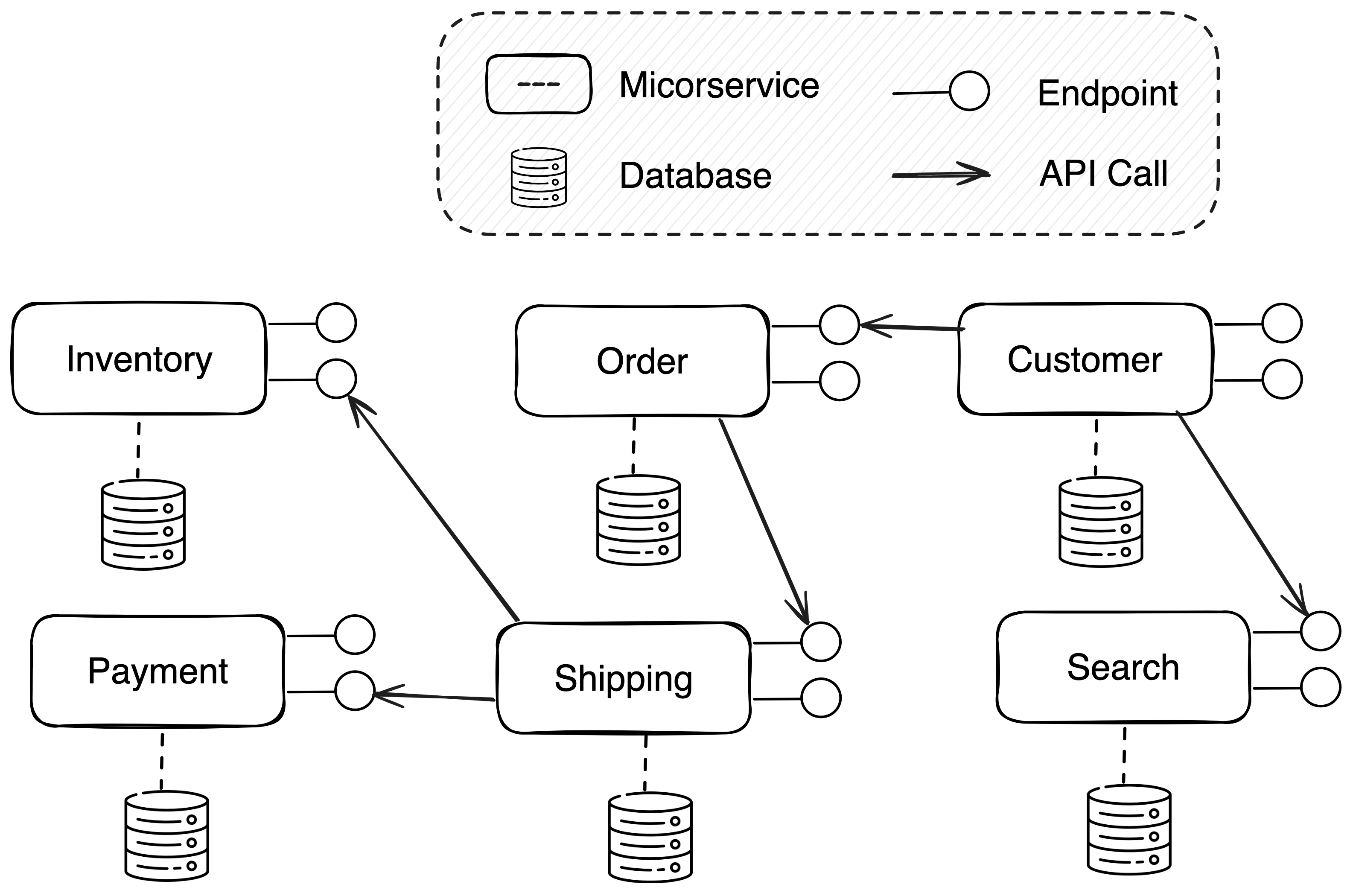}
\caption{Microservice Architecture of a Retail Application}
\label{fig:microservice_architecture}
\end{figure}

On the other hand, in a \textbf{microservice architecture}, as illustrated in Figure~\ref{fig:microservice_architecture}, each of these functionalities is treated as an independent microservice while the lines between them symbolize network communication, often using REST APIs. Each microservice is compiled and deployed separately and has its dedicated storage and resources. They communicate with each other over a network, often using REST APIs.

While both architectures strive for modularity, the representation of the decentralized nature of microservice architecture contrasts with the centralized, interconnected structure of modular architecture. However, the separation in microservice architecture enhances scalability and reduces complexity in managing individual components~\cite{springmicroservices}, however, it challenges the management of the holistic view of the system, while it distributes functionalities across independently deployable units.

\subsection{Frameworks for Modular Architecture}

Despite the recent surge in discussions surrounding module architecture, as revealed by a systematic grey literature review~\cite{su2024modular}, the results highlight the rising popularity of modular monoliths, which offer a fusion of monolithic and microservices benefits. The review encompassed 64 studies, with a significant increase observed in 2023. Alongside Service Weaver, two other frameworks listed by the authors for building modular monolith architecture are Spring Modulith and Light-hybrid-4j.

The Spring Modulith~\cite{spring-modulith} framework introduces an experimental approach for developing modular monolith applications within the Spring framework. It organizes source code based on the module concept, providing conventions and APIs for declaring and validating logical modules within Spring Boot applications. Key features include application modules representing units of functionality, module encapsulation, and restricted type visibility between modules. Additionally, it includes advanced features such as enhanced package arrangements, flexible module selection for integration tests, automatic developer documentation generation, runtime observability at the module level, and Passage of Time Events implementation.

The light-hybrid-4j~\cite{light-4j} framework, built upon the light-4j framework, is tailored for modular monolithic architectures and serverless deployments within the Light platform. It offers flexibility and cost savings in production by enabling modular deployment. The framework involves creating a server with integrated third-party dependencies and building multiple services with shared or specific dependencies. These services are compiled into independent jar files containing business handlers and loaded into a designated folder on the host server during startup. Traffic is routed to the appropriate handler in each service for incoming requests. Developers follow the principle of building services into jar files, deploying them to the same Docker container volume, and loading them into the same JVM. Services communicate through interfaces to conceal implementation details, and if necessary, services under heavy loads can be separated into individual containers for scalable deployment.

While still in development, Service Weaver~\cite{serviceweaver_hotos23} boasts superior capabilities compared to conventional modular frameworks. It is introudced to offer high performance, with co-located components communicating through direct method calls and remote components utilizing efficient custom serialization and RPC protocols. Additionally, Service Weaver requires minimal configuration and easily deploys to the cloud without extensive setup. It also provides libraries for logging, metrics, and tracing, seamlessly integrated into the deployed cloud environment.



\section{\uppercase{Examining Service Weaver through the Lens of Cloud-native Perspectives}}

Service Weaver packages core application logic in a modular binary structure for deployment, contrasting with the multiple codebases and deployments common in microservice architecture.
When evaluating a new technology or framework, it is customary to compare it to the prevailing standard practices. In the context of examining the advantages and drawbacks of Service Weaver, it proves advantageous to scrutinize each component and assess its functionality against the conventional implementation in a cloud-native system, adhering to established best practices, as outlined in the 12 factors~\cite{TheTwelv_10_3}.

The subsequent subsections iterate through the cloud-native components, exploring the offerings of Service Weaver in fulfilling each of them as summarized in~Table~\ref{tab:service-weaver-components}.





\begin{table}[h!]
\centering
\scriptsize
\setlength{\tabcolsep}{2pt}
\caption{Service Weaver Components}
\label{tab:service-weaver-components}
\begin{tabular}{p{4cm}p{1.2cm}p{1.5cm}}
 \textbf{Cloud-native Component}   & \textbf{Integrated} & \textbf{Dev. Respon.\textsuperscript{*}} \\
(C1) Services       & $\checkmark$ &     \\
(C2) Inter-Service Communication &  $\checkmark$ &            \\
(C3) Configuration Management & $\checkmark$ &   \\
(C4) Deployment        &   $\checkmark$ &          \\
(C5) Resiliency & & $\checkmark$ \\
(C6) Security    & & $\checkmark$     \\
(C7) Storage & $\checkmark$ &
\\
(C8) Logging & $\checkmark$ &
\\
(C9) Tracing & $\checkmark$ &
\\
(C10) Metric Tracking & $\checkmark$ &  \\
\end{tabular}
\scriptsize{Dev. Respon.\textsuperscript{*}: Developer Responsibility.}
\end{table}



\noindent\textbf{(C1) Services.}
Service Weaver’s main abstraction that functions in the place of a service is that of a component. A component is essentially a method implemented by Service Weaver that acts as a stand-alone software agent, handling its business logic and communicating with other components. Although polyglot support stands as a significant advantage for microservices and cloud-native systems, it's worth noting that Service Weaver currently only supports the GO programming language for implementation.

Moreover, the idea remains the same as a service, a component will have a narrow scope and will be stateless in its functionality. The difference is that all Service Weaver applications’ components are intended to be stored together in a single binary, whereas with microservices, services are not packaged within a single binary or else the system would be classified as a monolith.

By using a single binary with components, the code base could be more manageable for a single individual/group; especially if much of the boilerplate code has been abstracted away using Service Weaver. A problem that could arise with this single binary scheme is if an application grows to the point of needing to be managed by multiple teams. With a cloud-native system, each service has its code base and can be managed by a separate team without ever needing to cross paths with other services, teams, and their code bases. This tradeoff between centrally managed components and individually managed services needs to be considered when selecting an application’s architecture~\cite{serviceweaver_hotos23}.



\noindent\textbf{(C2) Inter-Service Communication.}
When compared to inter-service communications between Cloud-native services, inter-component communications using Service Weaver are much easier to digest and could lead to performance increases; however, they might lack some functionality. The way components communicate with each other using Service Weaver is as simple as calling a method. Since components are in the same binary, one component need only call another’s method that has been defined using an interface. Message passing happens locally for co-located components, otherwise, remote procedure calls are made for separate deployments. The abstraction of implementing remote procedure calls simplifies inter-component communication for developers, expediting the setup process and potentially enhancing performance, especially for local calls, where RPCs can incur significant overhead costs.



Digging into method calls being used for inter-component communication, it should also be noted that this abstraction would also abstract away the implementation details of service discovery and API gateway services. Service Weaver manages how it implements remote procedure calls and this includes knowing the deployed location of the components and load balancing the method calls appropriately. This means that instead of incurring all the heavy overhead of using something like Eureka to manage service discovery and then spinning up an API gateway service, a developer can simply call a component’s method and not have to worry about the semantics of how things get accomplished. This could greatly reduce development time and complexity.

While this abstraction has many benefits, it also restricts some functionality of a system that is implementing Service Weaver. To begin with, it looks like specific implementations of message passing, like priority queues, have yet to be implemented, though it does look like Service Weaver incorporates specific port listeners. Service Weaver documentation~\cite{serviceweaverblog} does not mention support for pre or post filters on messages, or manipulation of message headers, common features in API gateways for microservice architecture. One could argue that the input/output structure of the method being called could just be changed to accommodate for this; however, there could be bloat associated with the message passing in this way. Inter-component communications require balancing developer ease with control over functionality, influencing the choice between Service Weaver and traditional Cloud-native architecture.




\noindent\textbf{(C3) Configuration Management.}
In a Cloud-native system, there are usually many configuration files; typically, at least one per service, and sometimes this number can be exceeded if different environments are being used for each service. These config files are usually housed in some centrally available repo that is stored separately from the source code of the services. Configuration management can get cumbersome quickly and for this reason, Service Weaver aims to reduce configuration code as possible.

Service Weaver boasts a minimal configuration setup, as the libraries handle much of the components’ configurations automatically. This is yet another way in which Service Weaver aims to make life easier for the developer. There doesn’t need to be a central repo housing configuration files in this case, since everything is created and deployed with a single binary and there is also plenty of added functionality to the minimal config files, like the ability to declare ports and environmental variables.


\noindent\textbf{(C4) Deployment.}
For many development teams, one of the most challenging aspects of managing a project is the deployment and maintenance phases. Often there are dedicated teams solely responsible for this task called the DevOps team. A typical deployment for a Cloud-native system would be to package services together in containers using something like Docker, orchestrate those containers using tools like Kubernetes, and then finally deploy the services to some cloud source like Google Cloud, AWS, Azure, etc. All these actions take time, money, and a specified skill set.

With Service Weaver, component groupings are specified in the minimal configuration file, and then a simple call can be made from the command line to deploy all the components. At the time of writing, Service Weaver can deploy locally, deploy to a generic Kubernetes destination, or deploy to Google Kubernetes Engine (GKE) just by varying the command line prompt. This offers huge benefits in reducing the time and complexity of deploying an application. The fact that only one line must be written to transition from local development and testing to pushing an application to a Cloud is very impressive.

There is also the optionality to group components together while deploying. For instance, if one wanted to have component A and B running on a singles machine and have component C running on a separate machine, they once again can just add a few lines into the minimal configuration file to achieve this. This gives the developers/DevOps teams greater flexibility in choosing which components to group locally for an increase in performance.

While the ease of deployment is readily discussed within the Service Weaver documentation~\cite{serviceweaver_dev}, finer details about deployment implementations are necessarily available. Further experimentation should be conducted to understand if things like health checks and custom scripts can be implemented during the deployment of a Service Weaver implemented application. With this being considered, it is discussed that the Kubernetes deployer is easily integrated with common Continuous Integration/Continuous Deployment (CI/CD) pipelines like Github Actions, ArgoCD, and Jenkins, so probably much of the missing functionality could be implemented there.



\noindent\textbf{(C5) Resiliency.}
While microservices typically employ resiliency patterns like circuit breakers and fallbacks to enhance fault tolerance in client calls, Service Weaver diverges in its approach~\cite{aws_cloud_native}. As a modular monolith, Service Weaver doesn't engage in traditional client calls to other services but rather makes remote method calls, which may exhibit partial failures, participant failure, or result in errors. The documentation lacks specific details on methodologies or libraries for handling such errors, emphasizing that it is the responsibility of developers to detect and react to these issues~\cite{serviceweaver_hotos23,serviceweaver_dev}.


Consequently, developers are required to proactively address potential challenges such as timeouts or fallbacks in their code to ensure the system's resiliency. Despite this shift of responsibility to developers, it's noteworthy that the likelihood of latency and failures in method calls is considerably lower compared to service calls between microservices. The reduced likelihood of latency and failures in method calls in Service Weaver is due to its encapsulation of components within a singular binary. This enables shared memory space, promoting faster local method calls with minimized latency. The framework's emphasis on remote method calls adds control and predictability, diminishing the probability of failures. In contrast to microservices, where network-based inter-service communication may introduce latency and points of failure, Service Weaver's strategy optimizes communication within the same binary, resulting in a more reliable system with fewer failures.

Moreover, Service Weaver emphasizes the simplicity of conducting end-to-end tests, as applications are composed as single binaries in a unified programming language, making end-to-end tests akin to straightforward unit tests. This enables automated fault tolerance testing, such as chaos testing~\cite{basiri2016chaos} and model checking~\cite{lamport1994temporal}. This stands in contrast to the challenges posed by testability in distributed microservice architecture components~\cite{abdelfattah2023end}.


\noindent\textbf{(C6) Security.}
The absence of information in the documentation, articles, or blogs about the security features of Service Weaver was disappointing. However, one aspect that instills confidence is the potential lack of necessity for gateway services or gatekeeper services due to its modular monolith architecture. I conjecture that the security mechanism might be implemented at the method level, particularly concerning remote method calls, to ensure the integrity and confidentiality of data exchanged between components as well as the consistency of role access control across components~\cite{abdelfattah2023towards}.

\noindent\textbf{(C7) Storage.}
In a blog~\cite{mediumserviceweaver}, Shiju Varghese mentioned that Weaver diverges from frameworks like Encore, a distributed application framework in Go that natively supports SQL databases for storage by using PostgreSQL database ~\cite{mediumserviceweaver}. By refraining from mandating native support, Service Weaver allows users the flexibility to choose their preferred database, aligning with the characteristic flexibility of microservices. The storage functionality in Service Weaver additionally facilitates the seamless integration of storage-related preferences directly from configuration files, aligning with the microservices principle of separating configuration concerns from business logic. 


\noindent\textbf{(C8) Logging.}
Service Weaver employs the \textit{weaver.Logger} API to centralize logs originating from all components within the Service Weaver application. It offers the capability to seamlessly integrate logs with any cloud platform to which the application is deployed. This achievement significantly enhances developers' flexibility and improves the application's overall portability. Building upon this, the recognition of Service Weaver as a modular monolith further simplifies the process, eliminating the need for additional layers like log forwarding to platforms such as Logstash for log aggregation and tracing. This is a noteworthy advantage for Service Weaver, streamlining the logging process and contributing to its overall efficiency. 

\noindent\textbf{(C9) Tracing.}
Service Weaver seamlessly incorporates OpenTelemetry for tracing, providing a standardized method to collect and export traces effortlessly. Any trace generated within Service Weaver is automatically exported to the deployment environment, simplifying the developer's workflow. The platform boasts the capability to trace HTTP requests every second. Once tracing is activated, the system can autonomously trace both the HTTP request and the ensuing component method calls.~\cite{serviceweaver_hotos23,serviceweaver_dev}.

In contrast, in the microservices paradigm, developers have the flexibility to select from a range of tracing libraries such as Zipkin, Jaeger, and OpenTelemetry, tailoring their choice to the specific needs of the application. Unlike Service Weaver, automatic trace export is not built-in, posing challenges like context propagation and aggregation for developers.

\noindent\textbf{(C10) Metric Tracking.}
Service Weaver comes equipped with a built-in API catering to metrics like counters, gauges, and histograms. Its integration with deployment environments allows for the automatic export of metrics, ensuring a seamless connection with various platforms. In contrast, microservices offer a range of metrics options such as Prometheus and Grafana, providing developers with flexibility to choose tools that align with their service requirements. However, a notable distinction is that, unlike Service Weaver, microservices entail additional manual setup for metric services, making Service Weaver a more automatic and streamlined option~\cite{serviceweaver_hotos23,serviceweaver_dev}. 



\section{\uppercase{Discussion and Research Questions Answers}}



This section addresses the research questions posed in this paper, providing insights into the capabilities of Service Weaver and its preparedness for the development of a cloud-native system.


\vspace{-0.5em}
\subsubsection*{RQ$_{1}$: What advantages does writing your application as a modular binary offer in comparison to using Microservices?}
\vspace{-0.5em}
Service Weaver's modular binary approach, as opposed to traditional microservices architectures, brings forth a distinctive set of advantages. The framework's main abstraction, the component, operates like a stand-alone software agent, encapsulating business logic and communication functionalities. Unlike microservices, where services are dispersed and independently managed, Service Weaver components are stored together in a single binary, promoting enhanced code manageability, especially with significant abstraction of boilerplate code. 

\textit{Inter-component communications} in Service Weaver leverage method calls, simplifying the process and potentially enhancing performance. With components residing in the same binary, communication involves calling another component's method, eliminating the need for extensive service discovery and API gateway services. While this simplicity is beneficial, it may limit certain functionalities typically found in microservices architectures. For \textit{Configuration management} in Service Weaver is streamlined, requiring minimal centralized repositories for configuration files. The framework's libraries handle components' configurations automatically, reducing developer burdens and enhancing overall simplicity. \textit{Deployment and CI/CD} in Service Weaver are streamlined, emphasizing a single binary deployment. The framework minimizes configuration complexities, providing a straightforward deployment experience for developers. 

Moreover, \textit{Resiliency} in Service Weaver differs from microservices, relying on remote method calls that may exhibit partial failures. While microservices commonly implement resiliency patterns, Service Weaver shifts the responsibility to developers, but with a lower likelihood of latency and failures in method calls. Regarding \textit{Security} features in Service Weaver, while not extensively documented, suggest a potential lack of necessity for gateway services due to its modular monolith architecture. Security mechanisms may be implemented at the method level, warranting further investigation. 

Addtionally, \textit{Storage} in Service Weaver offers flexibility by not imposing native support, allowing developers to choose their preferred databases. Storage-related preferences are seamlessly integrated from configuration files, aligning with microservices' flexibility. \textit{Logging, metric tracking, and tracing} in Service Weaver are streamlined with built-in APIs and automatic exports to various platforms, contrasting with microservices where additional manual setup is often required. 

\vspace{-0.5em}
\subsubsection*{RQ$_{2}$: Is the Service Weaver capable of 
effectively covering cloud-native 
components?}
\vspace{-0.5em}
Service Weaver demonstrates commendable alignment with various cloud-native principles, encapsulating its functionalities in modular components stored together in a single binary. This approach enhances code manageability, although potential challenges may arise in the context of larger teams managing extensive applications. The framework streamlines configuration management by automatically handling component configurations, eliminating the need for extensive centralized repositories. Its deployment and CI/CD processes, featuring a single-binary deployment and integration with Kubernetes, significantly reduce deployment complexities. The framework's unique approach to inter-component communication simplifies the process through method calls, potentially leading to enhanced performance. Service Weaver supports logging, metric tracking, and tracing, showcasing its commitment to streamlined observability. 

It is crucial to highlight, that Service Weaver's approach may involve certain compromises. Offers less in the way of resilience and security solutions, and needs further examination. These trade-offs underscore the importance of balancing developmental ease with the need for flexibility and control in complex systems. On the other hand, the modular monolith architecture suggests potential security implementations at the method level.

Moreover, the absence of support for polyglot systems represents a notable disadvantage when Service Weaver is compared with microservice architecture. Microservice architecture, in contrast, offers the flexibility to employ heterogeneous languages and technologies within the same system.

Furthermore, considering the pivotal role of migration in this context, it's crucial to note that Service Weaver currently lacks a mechanism for migrating a microservices-based system to a Service Weaver-based system, whether through automated or semi-automated approaches. This limitation positions Service Weaver as a more suitable choice for systems that have not yet been implemented.



In summary, our examination of Service Weaver reveals its comprehensive support for diverse cloud-native features and components, such as modular services, inter-module communication, configuration management, CI/CD, flexible storage options, as well as logging, tracking, and tracing functionalities. While the framework adopts an approach of assigning resiliency and security responsibilities to developers, it exhibits an effective capability to handle cloud-native components. Nevertheless, it is worth noting that in its early release, the framework imposes limitations on the implementation of polyglot systems, while it only supports GO programming language.

\section{\uppercase{Conclusion}}


Service Weaver proposes a shift towards simplification and integration, aligning well with cloud-native principles like scalability and flexibility. The common theme when comparing the functionality of Service Weaver to that of a Cloud-native microservice system is that Service Weaver aims to make development and deployment as simple and seamless as possible for the development teams. Service Weaver uses many abstractions with their components and their communications to achieve this and they also implement many desired plugins like logging, tracing, and metric tracking out of the box. 

While there are compelling reasons for a development team to contemplate the adoption of Service Weaver, the ease of use it offers is accompanied by a few constraints. Currently, Service Weaver is exclusively implemented in GO, thus forfeiting the advantage of microservice architecture that accommodates polyglot systems. Additionally, the extensive use of abstractions in Service Weaver implies a trade-off in terms of total control over the system. Furthermore, as a relatively new project, there are certain functionalities that are yet to be developed, reflecting the ongoing maturation of the software. Notably, Service Weaver has not presented robust solutions for supporting system resiliency and security at this stage.



In the next research phase, we will implement a prototype GO application using the Service Weaver framework to assess development limitations in real-world scenarios.

\vspace{-0.5cm}
\section*{Acknowledgements}
\vspace{-0.2cm}

This material is based upon work supported by the National Science Foundation under Grant No. 2409933.






\bibliographystyle{apalike}
\bibliography{references}












\end{document}